\documentclass[onecolumn,prb]{revtex4}
\usepackage{epsfig,amssymb,amsmath}
\usepackage[english]{babel}
\begin{document}
\title{Determination of Cooper pairs and Majorana fermions currents ratio in DC-SQUID with topologically nontrivial barriers}
\author{I.~R.~Rahmonov$^{a,b}$, Yu.~M.~Shukrinov~$^{a,c}$, R.~Dawood~$^{d}$ and H.~El~Samman~$^{e}$}

%\author{Yu. M. Shukrinov~$^{1,2}$}
%\author{I. R. Rahmonov~$^{1,3}$}

\address{$^{a}$~Joint Institute for Nuclear Research, Dubna, Russia\\
$^{b}$~Umarov Physical and Technical  Institute, Dushanbe, Tajikistan\\
$^{c}$~Dubna State University, Dubna, Russia\\
$^{d}$~Cairo University, Giza, Egypt\\
$^{e}$~Menoufia University, Egypt
}

%\address{$^{1}$ BLTP, JINR, Dubna, Moscow Region, 141980, Russia \\
%$^{2}$ Dubna State University, Dubna,  141980, Russia\\
%$^{3}$Umarov Physical Technical Institute, TAS, Dushanbe, 734063 Tajikistan
%}

\date{\today}

\begin{abstract}
We present the results of numerical study of the phase dynamics of the DC-SQUID with topologically trivial and nontrivial barriers. In our calculations we take into account two components of supperconducting current, Cooper pairs ($2\pi$--periodic) and Majorana fermions ($4\pi$--periodic) currents. Magnetic field dependence of return current is presented. The qualitative behavior of this dependence is explained. We show that in case of two component superconducting current the periodicity of magnetic field dependence
 of return current displaced by Cooper pairs and Majorana fermion ratio over the magnetic field.
 This effect makes possible the experimental determination of ratio of Cooper pairs and Majorana fermions currents.
\end{abstract}

\maketitle

%\ead{rahmonov@theor.jinr.ru, ilhom-tj@inbox.ru}


%\maketitle
\section*{Introduction}
The Majorana fermions~\cite{majorana37} are attracting considerable interest in large part because of their application in quantum computers. Majorana fermions are predicted to exist in Josephson junctions with topologically nontrivial barriers~\cite{fu08,tanaka09}, where nontrivial states are formed. It is assumed that the nontrivial states are formed on the boundary or surface of a topological insulator~\cite{fu07} and a semiconductor nanowire in the presence of the Rashba spin-orbit coupling and the Zeeman field~\cite{sau10}.

The DC-SQUIDs with the nontrivial barriers are expected to be used as a real quantum gates~\cite{fu09}, since Majorana fermions exhibit a non-Abelian statistics which leads to a topological protection to errors. The formation of Majorana states in Josephson junction leads to tunneling of quasiparticles with charge $e$ in compare with $2e$ in usual case~\cite{kitaev01, veldhorst12}.
As a result, the oscillation period of the Josephson current is doubled $I_{s}=I_{c}\sin{\varphi/2}$~\cite{fu09}.
This $4\pi$--periodicity was discussed by A. Yu. Kitaev in Ref.~\cite{kitaev01}, where an experimental observation of Majorana fermions was suggested by the investigation of quantum wire bridge between two superconductors. Later Kwon at. el.~\cite{kwon04} have demonstrated a fractional ac Josephson effect, which is confirmation of $4\pi$ periodicity. The tunneling conductance peak at zero voltage was observed experimentally for the first time in the superconductor -- semiconductor nanowire junction~\cite{mourik12}, which hosts the Majorana fermions. An optimization study for Majorana fermions in a DC-SQUID with topologically nontrivial barriers was performed by Veldhorst et al.~\cite{veldhorst12}, where it was shown that the choice of the SQUID's parameters $\beta_L$ and $\beta_c$ can change the ratio of Majorana tunneling to a standard tunneling of Cooper pairs by more than two orders of magnitude.

In a recent work~\cite{rahmonov-jetpl16} we have shown that in the absence of external magnetic field value of return current for nontrivial case is in $\sqrt{2}$ time larger than its value in trivial case. Like the critical current, the return current of IV--curve of SQUID depends on external magnetic field. Additionally, the return current also depends on resonance features of SQUID. On the other hand, Majorana fermions lead to the changing of resonance feature of SQUID (due to $4\pi$-periodicity) and this question was discussed in Ref.~\cite{rahmonov-jetpl16}. It was shown that for the DC-SQUID with topologically nontrivial barriers the resonance branch~\cite{schmidt85} of IV--curve shifts by the $\sqrt{2}$ in comparison of trivial case. Thus, we expect that the investigation of the return current can be used as the tool for the study of Majorana fermions also. For such system the Josephson current consists of two components: Cooper pairs and Majorana fermions~\cite{dominguez12}. In this case one of the interesting question is a determination of the ratio of Cooper pairs and Majorana fermions currents.

In this work the results of numerical investigations of the phase dynamics of DC-SQUIDs are presented. Simulations are performed for the SQUID with the topologically trivial and nontrivial barriers, and for the case where Josephson current takes into account both components. The analysis of calculated current-voltage characteristics for the DC-SQUIDs are carried out. Also, we study the magnetic field dependence of return current for different parameters of the model, and demonstrate a possibility of the determination of ratio of Cooper pairs and Majorana fermions.

\section{Theoretical model and formulation}

Let us consider DC--SQUID with topologically nontrivial barriers. The presence of Majorana fermions leads to the single electron tunneling and doubles the period of phase difference of the order parameter. Therefore, within a resistively capacitively shunted junction (RCSJ) model (taking into account the existence of Majorana fermions) is sufficient to replace $2e$ by $e$ and $\varphi$ by $\varphi/2$ in the corresponding term of the system of equations.
Thus, for both cases, the Josephson relation in the nontrivial case is not changed, i.e.,
\begin{equation}
\label{Jos_rel}
\frac{\hbar}{e}\frac{d( \varphi/2)}{dt}=\frac{\hbar}{2e}\frac{d \varphi}{dt}=V
\end{equation}

\noindent where $\varphi$ and $V$ are the phase difference and voltage across JJ, respectively. The sum of currents for each JJ of DC-SQUID can be written as the following

\begin{equation}
\label{system_eq1}
\displaystyle I_{1,2}=\frac{C\hbar}{2e}\frac{\partial^{2}\varphi_{1,2}}{\partial t^{2}}+\frac{\hbar}{2eR}\frac{\partial \varphi_{1,2}}{\partial t}+ I_{c}\big[\alpha \sin\varphi_{1,2} +(1-\alpha)\sin\frac{\varphi_{1,2}}{2}\big]
\end{equation}

\noindent where $C$ is a capacitance, $R$ is a resistance and $I_{c}$ is a critical current of JJ, $\alpha$ is the ratio of the Cooper pairs and Majorana fermions currents through the junction, and $I_{1,2}$ are the currents through JJs of DC-SQUID and their sum equal to the external current $I=I_{1}+I_{2}$. We note that the ratio parameter $\alpha$ has been introduced first by Veldhorst with coauthors~\cite{veldhorst12}. We have modified the model by introducing of ratio parameter $\alpha$ for the expression of magnetic flux, while the authors of Ref\cite{veldhorst12} introduced it only for the expression of superconducting current. In the presence of the external magnetic field, the magnetic flux through circuit is quantized
\begin{equation}
\alpha(\varphi_{1}-\varphi_{2})+ (1-\alpha)\frac{\varphi_{1}-\varphi_{2}}{2} + \frac{\Phi_{t}}{\Phi_{0}}=2 \pi n
\label{flux_quantum}
\end{equation}
where $\Phi_{0}=h/2e$ is the magnetic flux quantum. The total flux $\Phi_{t}$ through DC-SQUID is determined by the expression
\begin{equation}
\label{total_flux}
\displaystyle \Phi_t=\Phi_{e}+L I_{c}\big[\alpha\sin\varphi_{1} +(1-\alpha)\sin\frac{\varphi_{1}}{2}- \big(\alpha\sin\varphi_{2} +(1-\alpha)\sin\frac{\varphi_{2}}{2}\big)\big]
\end{equation}
where $\Phi_{e}$ is the value of magnetic flux, created by the external magnetic field, $L$ is inductance of superconducting wires.

Using Josephson relation (\ref{Jos_rel}), expressions for currents (\ref{system_eq1}), magnetic field quantization condition (\ref{flux_quantum}), and expression for the total flux through DC-SQUID (\ref{total_flux}), we can write the system of equations that describes the dynamics of DC-SQUID in normalized units

\begin{equation}
\label{eq_2}
\left\{\begin{array}{ll}
\displaystyle \frac{\partial \varphi_{1,2}}{\partial t}=V_{1,2}
\vspace{0.3 cm}\\
\displaystyle \frac{\partial V_{1,2}}{\partial t}=\frac{1}{\beta_{c}}\bigg\{\frac{I}{2}
-V_{1}-\big[\alpha\sin\varphi_{1,2}+(1-\alpha)\sin\frac{\varphi_{1,2}}{2}\big]\\
\displaystyle \pm\frac{1}{2\beta_{L}}\bigg[2\pi( n-\varphi_{e})
\displaystyle -\big[\alpha(\varphi_{1} - \varphi_{2})+(1-\alpha)\frac{\varphi_{1} - \varphi_{2}}{2}\big]\bigg]\bigg\}
\end{array}\right.
\end{equation}

\noindent where $\beta_{c}=2 \pi I_c R^2 C /\Phi_0$ is a McCumber parameter, $\beta_L =2 \pi L I_c/ \Phi_0$ is normalized inductance, $\varphi_{e}=\Phi_{e}/\Phi_{0}$ is normalized external magnetic flux. In the system of equations (\ref{eq_2}), time is normalized to $\omega_{c}^{-1}$, where $\omega_{c}=2eI_{c}R/\hbar$, voltage -- to the $V_{c}=I_{c}R$ and bias current $I$ -- to the critical current $I_{c}$.

The capacitance of the Josephson junction and the inductance of the superconducting wires in the DC-SQUID form an oscillatory circuit with frequency $\omega_{r}$, determined with expression (\ref{res_frequency}). When the condition $\omega_{J}=m\omega_{r}$ ($m$ is integer number) is satisfied, the branches appear in the IV--curve, their origin is associated with the resonance of Josephson and electromagnetic oscillations~\cite{schmidt85}

\begin{equation}
\label{res_frequency}
\omega_{r}=1/\sqrt{\beta_c \beta_L}
\end{equation}

In our calculations we assume that $\beta_{c}=10$ and $\beta_{L}=1$. The Bias current increases from $I_{0}=0.1$ till $I_{max}=2.3$ and further reduces to zero by step of $\Delta I=0.0005$. At each fixed value of the bias current, the system of differential equations (\ref{eq_2}) are solved by the fourth order Runge-Kutta method in time interval from $0$ to $T_{max}=5000$ with a step $\Delta t=0.05$. As a result we have obtained a voltage $V$ and a phase difference $\varphi$ as a function of time. Then, the obtained values of the voltage $V$ is averaged in a time interval [50,5000].

\section{Results and discussions}

\begin{figure}[h!]
 \centering
 \includegraphics[width=55mm]{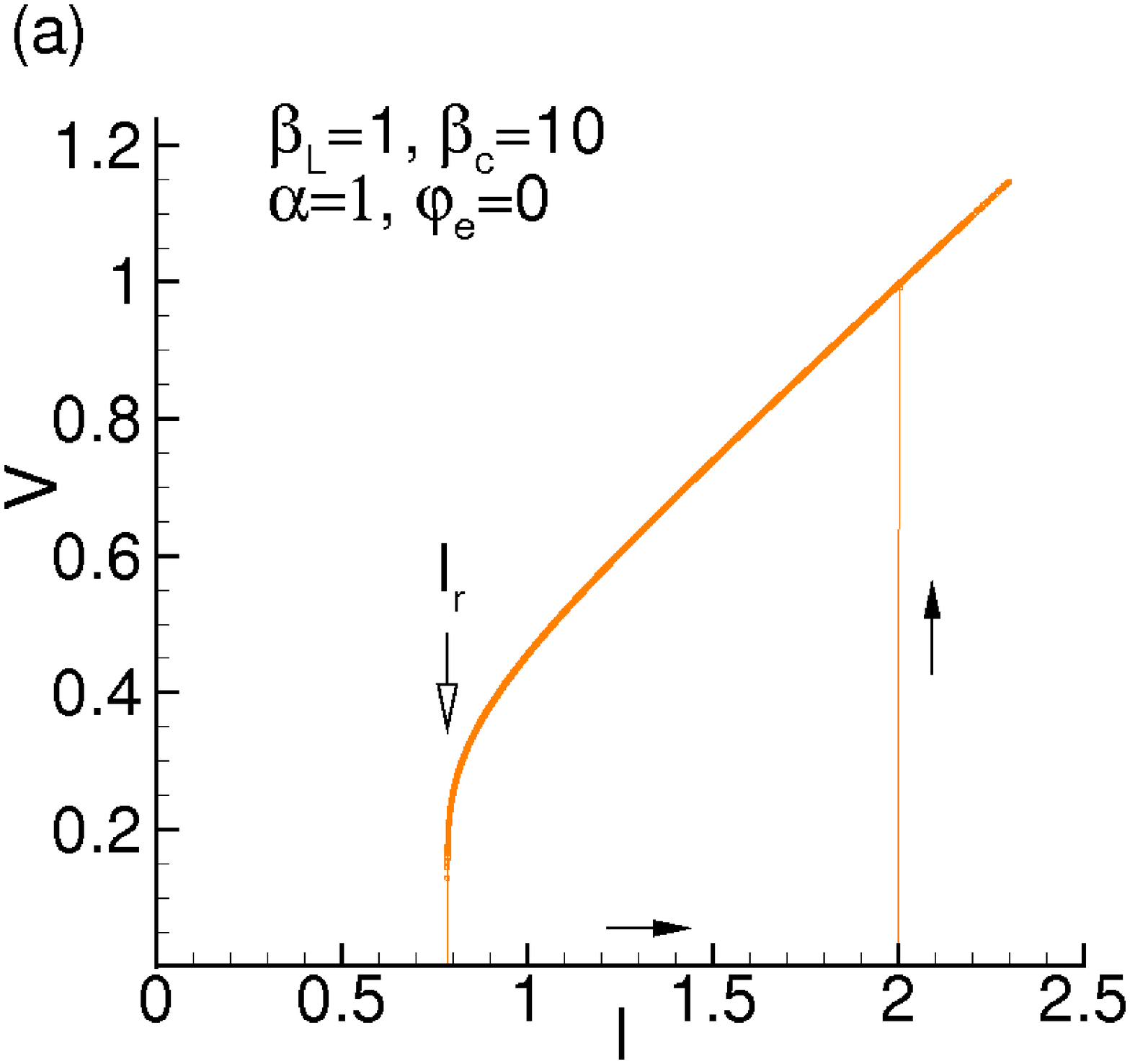}
 \includegraphics[width=55mm]{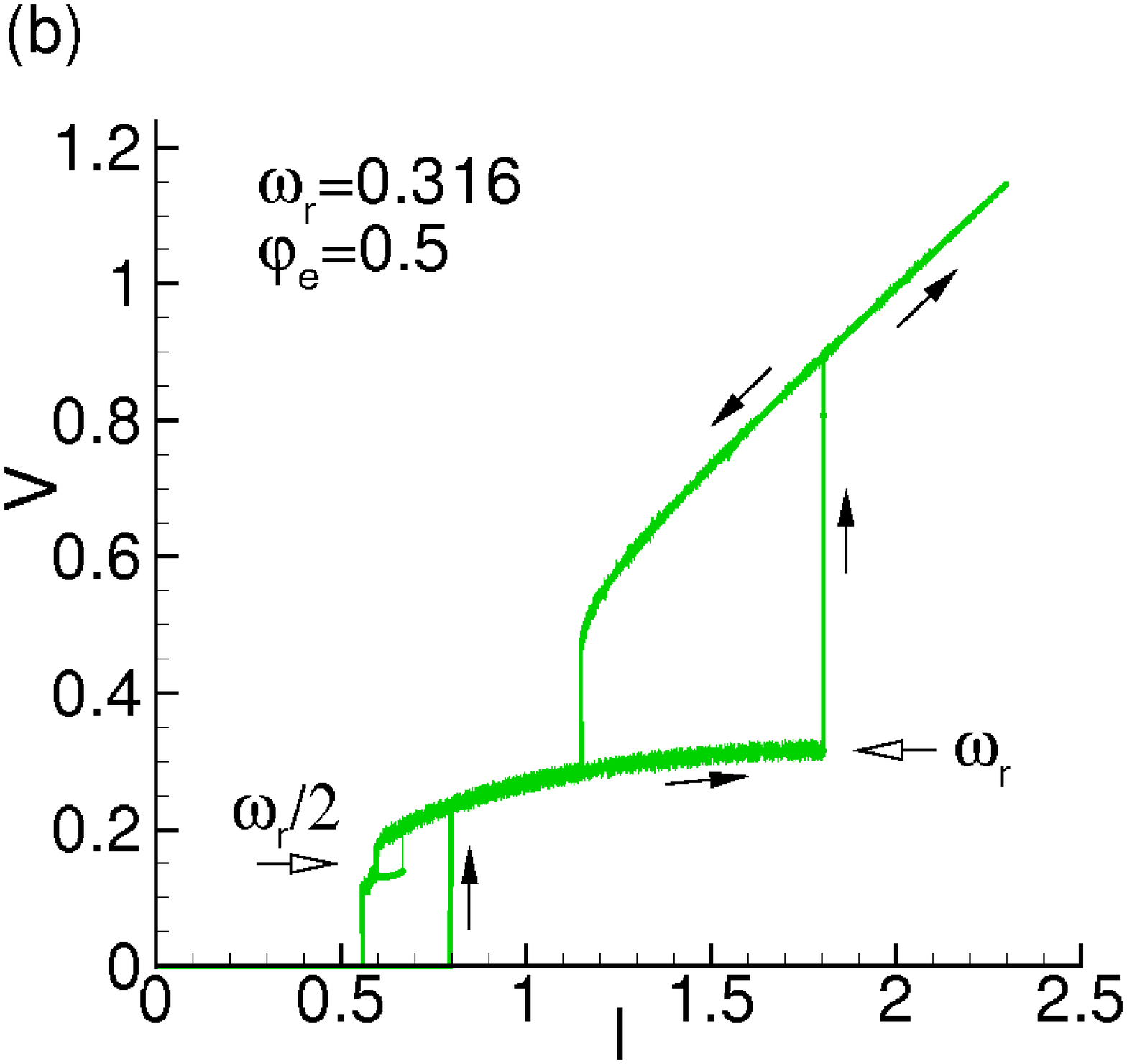}
\caption{(a) IV--curve for trivial case ($\alpha=1$) at $\varphi_{e}=0$. Filled arrows show the direction of the bias current sweeping. A hollow arrow shows the value of return current, which corresponds to the transition of JJs of SQUID from resistive ($V\neq0$) to the superconducting  ($V=0$) state. (b) The same as (a) at $\varphi_{e}=0.5$. The hollow arrows show the position of main resonance branch and its subharmonic.}
\label{cvc_alpha1_phe0_05}
\end{figure}

Let us first discuss the phase dynamics of the DC-SQUID with trivial barriers, i.e. the case with $\alpha=1$ in the equation (\ref{eq_2}). The IV-characteristic in the absence of an external magnetic field is presented in Fig.\ref{cvc_alpha1_phe0_05}(a). The obtained IV-characteristic demonstrates a hysteresis as it should be for the underdamped case ($\beta_{c}>1$). The value of return current in this case is equal to 0.7853.

The external magnetic field leads to the changing of the IV-characteristic. To demonstrate its effect, we have calculated the IV-characteristic for the DC-SQUID with trivial barriers at the value of magnetic field $\varphi_{e}=0.5$, which is shown in Fig.\ref{cvc_alpha1_phe0_05}(b). This characteristic shows two additional branches in comparison of IV--curve in Fig.\ref{cvc_alpha1_phe0_05}(a) caused by the resonance of Josephson and electromagnetic oscillations in the SQUID~\cite{schmidt85}. The upper branch corresponding to the frequency $\omega_{J}=\omega_{r}=0.316$ is the main resonance branch and another one corresponding to the $\omega_{J}=\omega_{r}/2=0.158$ is its subharmonic. We note that the Josephson frequency $\omega_{J}$ and voltages $V_{1,2}$ in JJs have the same value in our normalization. One can see that the transition from resistive to the superconducting state occurs through the subharmonic resonance branch corresponding to $\omega_{J}=\omega_{r}/2$. In this case the value of return current is equal to $I_{r}=0.5584$, which is smaller than in case $\varphi_{e}=0$. So, we mark the fact, that the external magnetic field leads to the changing of return current value.

\begin{figure}[h!]
 \centering
 \includegraphics[width=55mm]{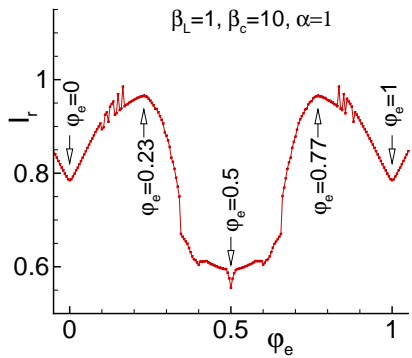}
\caption{External magnetic field dependence of return current $I_{r}(\varphi)$ for $\alpha=1$ (trivial case). Arrows mark the value of maximums and minimums.}
\label{Ir_phe_alpha1}
\end{figure}

In order to perform a detailed analysis of the influence of external magnetic field on the return current, we have calculated its dependence on magnetic field presented in Fig.\ref{Ir_phe_alpha1}. This dependence $I_{r}(\varphi_{e})$ is calculated for the values of $\varphi_{e}$ in the interval [0,1], because for other values it is the same. We stress here that in chosen interval the $I_{r}(\varphi_{e})$ demonstrates the minimums of return current at the values of magnetic field equal to $\varphi_{e}=0$, $\varphi_{e}=0.5$ and $\varphi_{e}=1$ and maximums at $\varphi_{e}=0.23$ and $\varphi_{e}=0.77$. The return currents at $\varphi_{e}=0$ and $\varphi_{e}=1$ are the same and equal to $I_{r}=0.7853$. The maximum of $I_{r}(\varphi_{e})$ at $\varphi_{e}=0.23$ and $\varphi_{e}=0.77$ is equal to $I_{r}=0.966$.

\begin{figure}[h!]
\centering
\includegraphics[width=55mm]{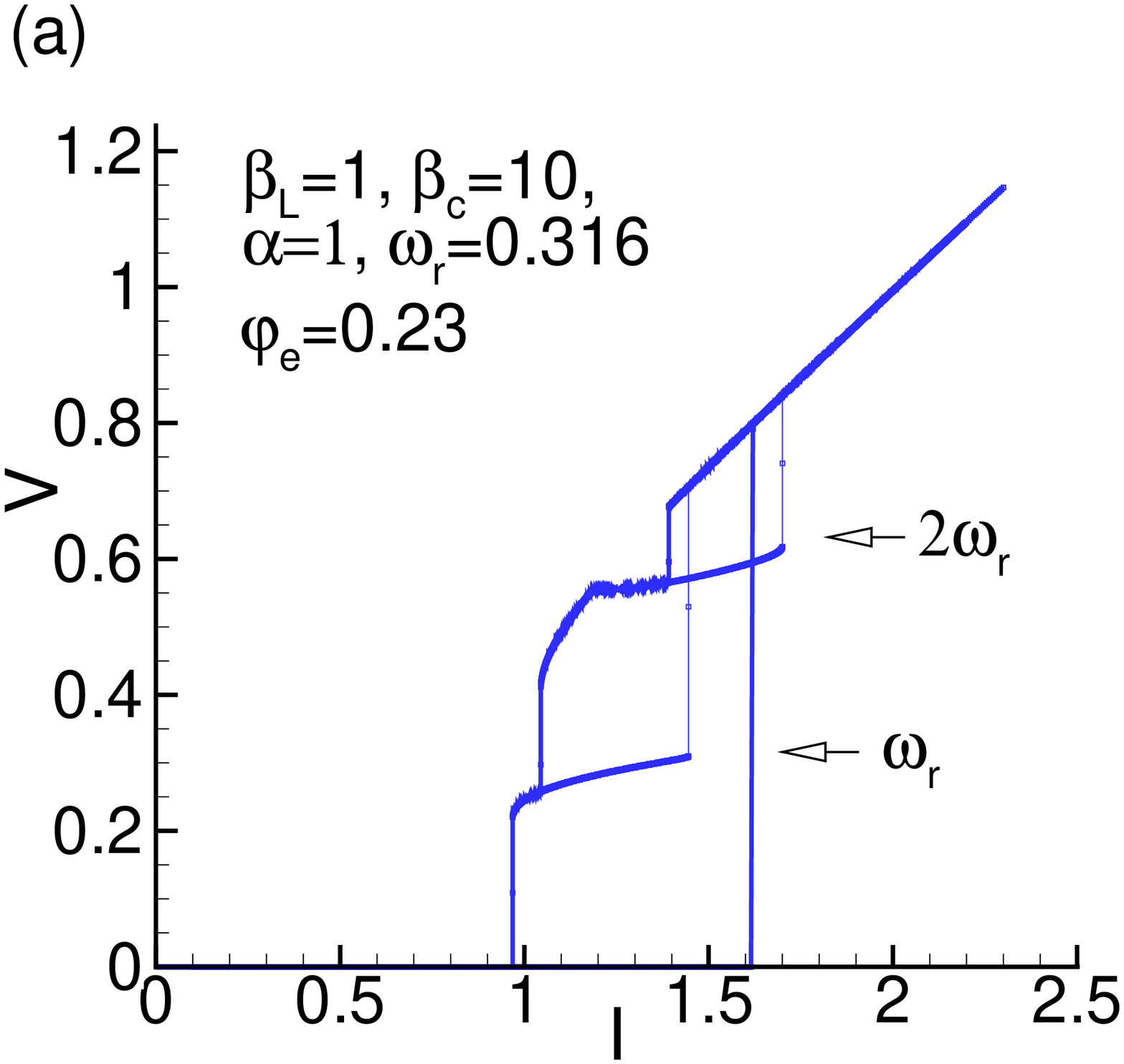}
\caption{Manifestation of the transition from resistive to the superconducting state in IV--curve for trivial case at $\varphi_{e}=0.23$. Arrows indicate the position of main resonance branch and its harmonic.}
\label{cvc_alpha1}
\end{figure}

The origin of the observed minimums and maximums can be explained by the analysis of IV--curves calculated for the corresponding values of magnetic field. Let us analyze the maximum of $I_{r}(\varphi_{e})$. The IV-curve for the value of the external magnetic field $\varphi_{e}=0.23$ corresponding to the maximum of $I_{r}(\varphi_{e})$ is presented in Fig.\ref{cvc_alpha1}. This curve demonstrates the main resonance branch and its harmonic corresponding to the frequencies $\omega_{J}=\omega_{r}=0.316$ and $\omega_{J}=2\omega_{r}=0.632$, respectively. One can see that the transition from the resistive to the superconducting state takes place through the main resonance branch. The same feature is obtained at $\varphi_{e}=0.77$, which corresponds to the maximum of $I_{r}(\varphi_{e})$. This fact allows to conclude that the maximum of $I_{r}(\varphi_{e})$ can be observed when the transition occurs through the main resonance branch corresponding to $\omega_{J}=\omega_{r}$.

Let us now analyze minimums of $I_{r}(\varphi_{e})$. In case of $\varphi_{e}=0$, the resonance branch does not occur (see Fig.\ref{cvc_alpha1_phe0_05}(a)), and the transition of JJs of SQUID from the resistive to the superconducting state takes place without resonance. The same result is observed at $\varphi_{e}=1$. So, the absence of the resonance  is a condition for the appearance of minimums at $\varphi_{e}=0$ and $\varphi_{e}=1$. We call such minimums as ``nonresonant minimums''. Another minimum of $I_{r}(\varphi_{e})$ caused by the transition of JJs from the resistive to the superconducting state through the subharmonic resonance branch corresponding to the frequency $\omega_{J}=\omega_{r}/2$. Its follows from the analysis of IV--curve at $\varphi_{e}=0.5$ (see Fig.\ref{cvc_alpha1}).

\begin{figure}[h!]
 \centering
 \includegraphics[width=55mm]{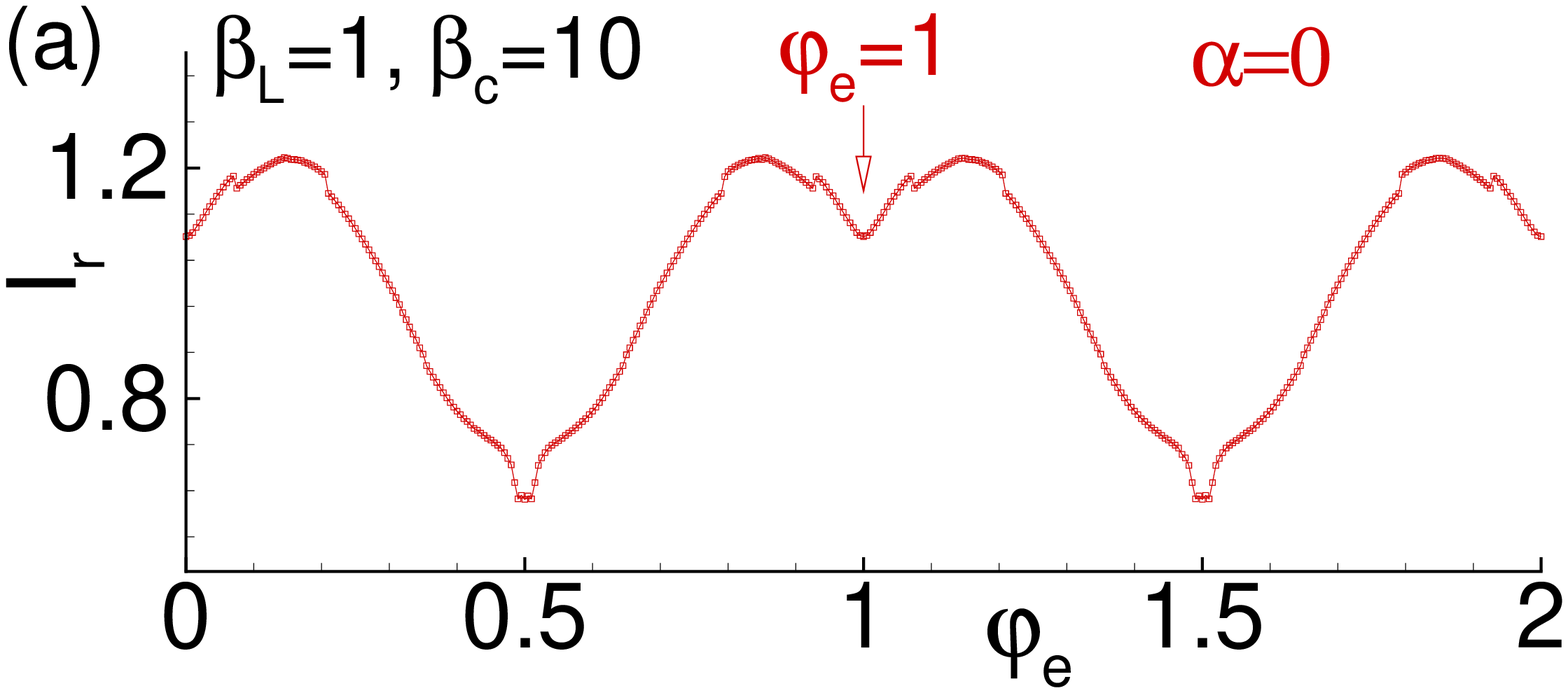}\\
 \includegraphics[width=55mm]{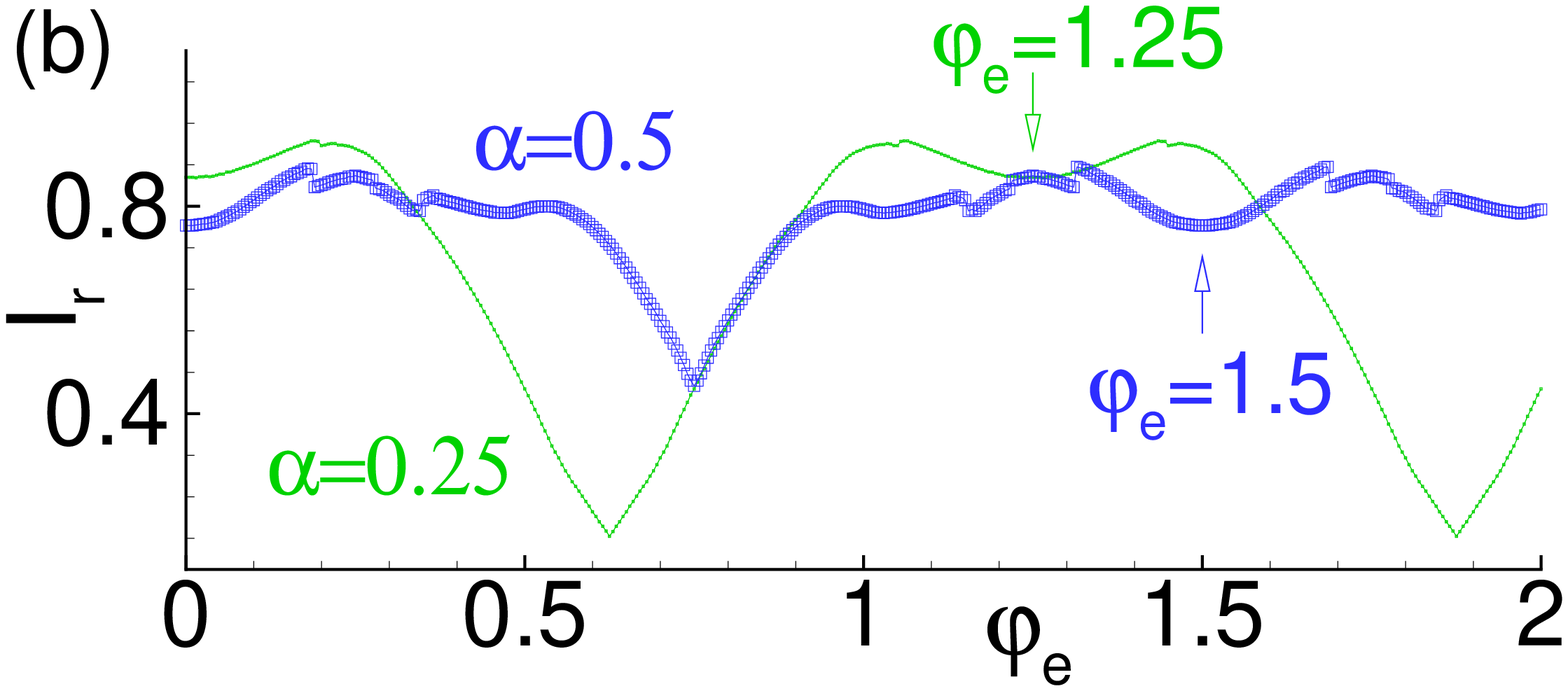}
\caption{(a) Field dependence of return current for $\alpha=0$ (nontrivial case). The position of nonresonant minimum at $\varphi_{e}=1$ is shown with the hollow vertical arrow. (b) The same as (a) for $\alpha=0.25$ and $\alpha=0.5$.}
\label{Ir_compare}
\end{figure}

Now we consider a nontrivial case, i.e. the case with $\alpha=0$. The external magnetic field dependence of $I_{r}(\varphi_{e})$ is presented in Fig.\ref{Ir_compare}(a). The periodicity of $I_{r}(\varphi_{e})$ coincides with  the trivial case, but for all values of $\varphi_{e}$, the return current $I_{r}$ is larger than in trivial case. Another difference concerns the fact that the return current in nontrivial case at the value of external magnetic field corresponding to the nonresonant minimum ($\varphi_{e}=0$, $\varphi_{e}=1$) is $\sqrt{2}$ time larger than in trivial case~\cite{rahmonov-jetpl16}.

Results for the intermediate values of $\alpha$, when superconducting current takes into account both component presented in Fig.\ref{Ir_compare}(b), which shows the dependence of return current $I_{r}$ on external magnetic field for the $\alpha=0.25$ (thin line) and $\alpha=0.5$ (thick line). Here the above mentioned nonresonant minimum appears at $\varphi_{e}=1.25$ for the case $\alpha=0.25$ and at $\varphi_{e}=1.5$ for the case $\alpha=0.5$. Positions of the nonresonant minimums are shown by the vertical arrows. So, one can see that in case of both component of superconducting current the nonresonant minimum shifts along $\varphi_{e}$ in comparison with a case $\alpha=0$ to the value, which corresponds $\alpha$ and the periodicity of $I_{r}(\varphi_{e})$ changes to the corresponding value of $\alpha$. With increasing of $\alpha$ the value of $I_{r}$ decreases and for $\alpha=1$ it decreases $\sqrt{2}$ times, which coincides with the results published in Ref.\cite{rahmonov-jetpl16}.

\begin{figure}[h!]
 \centering
 \includegraphics[width=60mm]{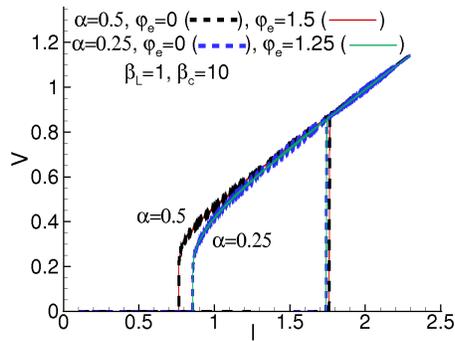}
\caption{Demonstration of the coincidence of IV--curves at $\varphi_{e}=0$ (dashed blue online) and $\varphi_{e}=1.25$ (solid green online) for $\alpha=0.25$ and at $\varphi_{e}=0$ (dashed black online) and $\varphi_{e}=1.5$ (solid red online) for $\alpha=0.5$.}
\label{cvc_alpha025_05}
\end{figure}

In order to demonstrate that the observed minimums are the nonresonant ones, we have calculated IV--curves for the values of external magnetic field corresponding to these minimums. Figure \ref{cvc_alpha025_05} shows them for $\alpha=0.25$ at $\varphi_{e}=1.25$ (solid line) and $\varphi_{e}=0$ (dashed line) and also the IV--curves for $\alpha=0.5$ at $\varphi_{e}=1.5$ (solid line) and $\varphi_{e}=0$ (dashed line). As we expected, the IV--curve at $\varphi_{e}=1.25$ absolutely coincides with the IV--curve at $\varphi_{e}=0$ and demonstrates that the transition from resistive state to the superconducting one takes place without a resonance. As it was mentioned above, it is a condition of nonresonant minimum. A similar behavior is also observed for the case with $\alpha=0.5$: these IV--curves also demonstrate the above mentioned transition without resonance branch, the $\varphi_{e}=1.5$ corresponds to the nonresonant minimum.

\section{Conclusion}
In conclusion, the peculiarities of the phase dynamics of DC--SQUID with trivial and nontrivial barriers
have been studied numerically. The detailed analysis of effect of the magnetic field on the return current $I_{r}(\varphi_{e})$ in the IV--characteristic of DC--SQUID has been carried out. We have shown that the maximum of magnetic field dependence of return current $I_{r}(\varphi_{e})$ corresponds to the case when transition of IV--curve from resistive to the superconducting state takes place through the main resonance branch. The performed analysis demonstrates two type of minimums in $I_{r}(\varphi_{e})$ dependence. One of them corresponds to the case when transition from resistive to the superconducting state takes place through the subharmonic resonance branch. Another one occurs when transition happens without resonance (nonresonant minimum). In case of two components of superconducting current, a nonresonant minimum shifts along the magnetic field to the value, which corresponds to the value of parameter $\alpha$ determining the ratio of the Cooper pairs and Majorana fermions currents. We assume that it may be used for the experimental determination of this currents ratio.

\section{Acknowledgements}
The reported study was funded by RFBR according to the research project 15-51-61011 and 16-52-45011

{}

\end{document}